\documentclass[
 reprint,
 amsmath,amssymb,
 aps,
]{revtex4-1}
\usepackage[colorlinks]{hyperref}

\usepackage{graphicx}
\usepackage{dcolumn}
\usepackage{bm}
\usepackage{blindtext}
\usepackage{tabularx}
\usepackage{pifont}
\usepackage{subfigure}
\usepackage{caption}
\usepackage{dsfont}
\usepackage{comment}
\usepackage{float}
\usepackage{mathtools}

\newcommand{\bra}[1]{\langle #1 |}
\newcommand{\ket}[1]{| #1 \rangle}

\newcommand{\tr}{\mathrm{Tr}}

\newcommand{\bea}{\begin{eqnarray}}
\newcommand{\eea}{\end{eqnarray}}
\newcommand{\be}{\begin{equation}}
\newcommand{\ee}{\end{equation}}
\newcommand{\rsub}[1]{_{\mbox{\tiny{#1}}}}

\newcommand{\nn}{\nonumber}
\newcommand{\xmark}{\text{\ding{55}}}
\DeclareMathOperator{\sech}{sech}
\graphicspath{{images/}}
\begin{document}
\title{Parity-Swap State Comparison Amplifier for Schr\"odinger Cat States} 
\author{Gioan Tatsi, Luca Mazzarella and John Jeffers}
\address{Department of Physics, University of Strathclyde, John Anderson Building, 107 Rottenrow, Glasgow G4 0NG, U.K.}
\date{\today}
\begin{abstract}
We propose a postselecting parity-swap amplifier for Schr\"odinger cat states that does not require the amplified state to be known a priori. The device is based on a previously-implemented state comparison amplifier for coherent states. It consumes only Gaussian resource states, which provides an advantage over some cat state amplifiers. It requires simple Geiger-mode photodetectors and works with high fidelity and approximately twofold gain.
\end{abstract}
\maketitle
\section{Introduction}

Schr\"odinger cat states, superpositions of two coherent states with coherent amplitudes of the same magnitude but different phases, have been widely studied for the significant role they could play in quantum information 
\cite{Cochrane, Nielsen, Ralph1, Gilchrist, Tipsmark}, computation \cite{Lund, Lobino} and in fundamental tests \cite{Jeong, Strobinska, Lee, McKeon} as resource states. For example Ralph \textit{et al.} \cite{Ralph1} showed that cat states can be used to implement qubit gates in an all optical quantum computation scheme,  where the logical qubits are encoded in the phase of the coherent state complex amplitude and Jeong \cite{Jeong} considered the possibility of testing the Bell inequalities using cat-like states as resources.

The nongaussian character of cat states renders them both challenging to engineer and fragile. Traditionally they have been generated using a combination of linear optics and measurement postselection techniques \cite{Gerrits, Wakui, Ourjoumtsev, Nielsen2, Takeoka, Marek, Lee2, Kim}. One simple implementation relies on subtracting a photon from a squeezed vacuum state. This generates a state close to an odd cat state, but only for small coherent amplitudes ($\alpha \leq 1.2$)\cite{Laghaout}. However, some quantum computation schemes require as resources cat states of higher coherent amplitude  \cite{Ralph1, Lund}. 
One way of increasing the amplitude to meet this need relies on the process of cat state ``breeding"\cite{Lund2, Takeoka2, Laghaout, Synchev}, whereby small amplitude ``kittens" are mixed coherently at a beam splitter and a postselection measurement projects one of the two outputs into an increased amplitude state based on the constructive interference of the two input modes. This can be done recursively and, in principle, it can produce cat states of arbitrary size. There is, however, an unfavourable scaling of both the output quality and the success probability with increased ``breeding seasons''. Also, the smaller amplitude nongaussian cat states are themselves consumed at each recursion to make larger amplitude states. Given that cat states are expensive quantum resources this is not a particularly desirable state of affairs.

  Recently, there has been a renewed interest in the amplification of cat states  \cite{Evgeny, Lidal}. Here we introduce an optical amplifier for Schr\"odinger cat states that works without requiring prior knowledge of the input state, that relies on Gaussian resources, beamsplitters and Geiger mode detectors and offers a reasonably high 
gain and high output quality for a range of low amplitude cat states.
The proposed scheme is inspired by the so called state comparison amplifier (SCAMP) \cite{Electra, Ross, Ross2, Ross3, David,Luca} a non-deterministic amplifier that has been shown to achieve high gain and high output quality for discrete sets of input coherent states of light. 

The article is organised as follows: in Section II we introduce the Schr\"odinger cat states, in particular the even and odd cats. Section III reviews the state comparison amplifier for coherent states. In Section IV we introduce the Parity-Swap Amplifier for cat states and we benchmark the performance of the scheme based on the fidelity and the success probability.

\section {Schr\"odinger Cat States}
Optical coherent states are superpositions of all photon numbers, 
 \bea
 \ket{\alpha} = e^{-|\alpha|^2/2}\sum_{n=0}^\infty \frac{\alpha^n}{\sqrt{n!}} \ket{n},
 \eea
where $\alpha$ is a complex number known as the coherent amplitude \cite{Glauber}. 
The states therefore have photon number probabilities that satisfy a Poisson distribution with mean $|\alpha|^2$. Different coherent states are not orthogonal and so cannot be distinguished perfectly  and deterministically by any measurement . They approach orthogonality if their coherent amplitudes are sufficiently far apart and the states can be distinguished in this limit. Coherent states of low mean photon number naturally do not have very different amplitudes and their indistinguishability is the only quantum property that can be measured. For this reason coherent states are often described as the most classical of quantum states.

Optical Schr\"odinger cat states are superpositions of two coherent states with coherent amplitudes of the same magnitude but different phases, typically opposite,
 \begin{align}
	\ket{\alpha_\theta}&=\frac{1}{\sqrt{ 2+2\cos{\theta}e^{-2|\alpha|^2}}}(\ket{+\alpha}+e^{i\theta} \ket{-\alpha}).
\end{align}
For high values of the mean photon number $|\alpha|^2$ cat states are superpositions of states that can be distinguished macroscopically. This is in analogy with Schr\"odinger's original gedanken experiment, which was introduced \cite{Schrodinger} in order to advocate the difficulties in applying quantum mechanics to the macroscopic world. 

Typically we set $\theta=\{0,\pi\}$, obtaining the so called {\itshape even} and {\itshape odd} cat states
 
  \begin{align}
	\ket{\alpha_\theta}&=\frac{1}{\sqrt{2 +2\cos{\theta} e^{-2|\alpha|^2}}}(\ket{\alpha}+e^{i\theta} \ket{-\alpha})\nonumber\\
	&=\frac{e^{\frac{-|\alpha|^2}{2}}}{\sqrt{2 +2\cos{\theta} e^{-2|\alpha|^2}}} \sum_{n=0}^\infty \frac{\alpha^n +e^{i\theta} (-\alpha)^n}{\sqrt{n!}} \ket{n} \nonumber \\
	&= \begin{cases}
            \frac{1}{\sqrt{\cosh |\alpha|^2}}\sum_{n=0}^\infty \frac{\alpha^{2n}}{\sqrt{(2n)!}} |2n \rangle & \theta=0 \\
            \frac{1}{\sqrt{\sinh |\alpha|^2}}\sum_{n=0}^\infty \frac{\alpha^{2n+1}}{\sqrt{(2n+1)!}} |2n+1 \rangle & \theta=\pi
        \end{cases},
\end{align}
so the even cat state $\ket{\alpha_0}=\ket{\alpha_+}$ contains only even photon numbers and conversely the odd state $\ket{\alpha_\pi}=\ket{\alpha_-}$ only has odd photon numbers. The even and odd states are mutually orthogonal and, despite the fact that they are superpostions of almost classical states, the gaps in the photon number distribution are a signature of nonclassicality, whatever the value of $|\alpha|$. They are also nongaussian, as attested by the negativity of their Wigner function, another widely accepted measure of nonclassicality \cite{Hillery}. 

\section{State Comparison Amplifier}
\begin{figure}[ht]
	\centering
	\includegraphics[trim={4cm 3.5cm 4cm 1.5cm},clip,width=9cm]{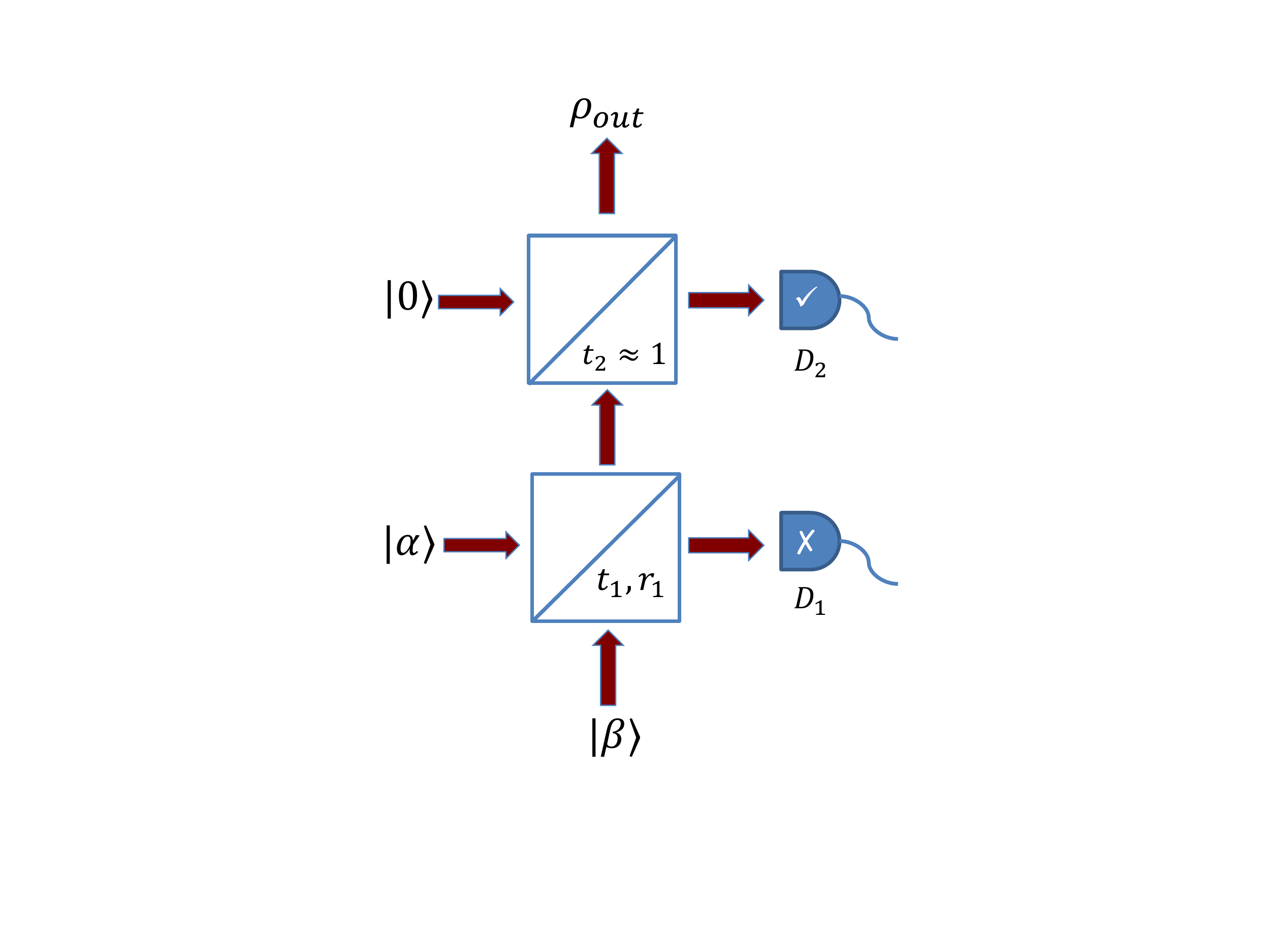}
	\caption{
	State Comparison Amplifier comprised of a comparison stage and a subtraction stage.}
	\label{scamp}
\end{figure}

 As previously mentioned, the state comparison amplifier (SCAMP) is a nondeterministic amplifier for coherent states that works with high gain, provides high-quality output and requires only classical resources. The backbone of the amplifier (Fig. \ref{scamp}) is the mature technique of state comparison \cite{Erika}, which has been used in the setting of multiple phase encoding quantum receivers \cite{Kennedy, Dolinar, Bondurant}. In our scheme the coherent state to be amplified is drawn uniformly at random from a discrete set, say $\{\ket {+\alpha}, \ket {-\alpha}\}$ and is mixed with a guess coherent state drawn from the set $\{\ket {+\beta}, \ket {-\beta}\}$ on a beamsplitter with transmission and reflection coefficients $t_1$ and $r_1$ respectively. Here we assume that the reflection and transmission coefficients are real, so the light incident on the lower input arm picks up a $\pi$ phase shift upon reflection. 
 The beamsplitter relates the output mode annihilation operators to those for the input mode as follows
 \begin{align}
     \begin{pmatrix}\hat{a}_{out}\\\hat{b}_{out}\end{pmatrix}=\hat{U}_{BS}\begin{pmatrix}\hat{a}_{in}\\\hat{b}_{in}\end{pmatrix}
 \end{align}
where 
\begin{equation}
    \hat{U}_{BS} =\begin{pmatrix}t&-r\\r&t\end{pmatrix}
   \end{equation}
is the  beamsplitter transformation matrix with real transmission and reflection coefficients. 

Then if the input states chosen are $|\alpha \rangle$ and $|\beta \rangle$ the state after the beamsplitter is
\begin{equation}
	\hat{U}_{BS}\ket{\alpha,\beta}\rightarrow\ket{t_1 \alpha-r_1 \beta,t_1\beta+r_1 \alpha}
\label{beamsplitter}
\end{equation}

From eq. \ref{beamsplitter} it is not difficult to see that the amplitude in the mode in the detector arm vanishes if $t_1 \alpha=r_1 \beta$, i.e when the guess state amplitude is $\beta=\frac{t_1\alpha}{r_1}$. The detector cannot fire. We choose this amplitude for $\beta$ and use the non-firing of the detector as an indication of the success of the amplification process and accept the output whenever it occurs.

If, however, the input and guess states are oppositely-phased, say $\alpha$ and $- \frac{t_1\alpha}{r_1}$, light leaks into the detector arm and the detector can fire, in which case the output state is rejected. The detector may not fire, however, because the coherent state is not orthogonal to the vacuum, $\langle \alpha|0 \rangle=\langle -\alpha|0 \rangle=e^{-\frac{|\alpha|^2}{2}}$. The detector not firing is therefore an {\itshape imperfect indication} that the correct guess state has been chosen. This can limit how well the output state mimics to the nominal amplified output. Conditioned on the detector not registering an event, the output of SCAMP after the comparison state is a mixture of the nominal correctly amplified state $|g \alpha \rangle =|\alpha/r_1\rangle$ and another, lower amplitude coherent state, with their respective probabilities biased towards the amplified state. 

 We use the quantum fidelity, $\mbox{F}$, of the output state to the nominal amplified state as a measure of the output quality \cite{Barnett2}. As one of the states is pure, here the fidelity is given by 
\begin{align}
	\mbox{F}= &\tr[|g \alpha \rangle \langle g\alpha|\rho\rsub{out}]\\
	= &\int_{\mathbb{C}} \frac{d^2\xi}{\pi} \chi_{g\alpha}(-\xi)\chi_{out}(+\xi)
\end{align}
where in the second line the fidelity is written in terms of the characteristic functions of the nominal and actual output states. The symmetrically ordered characteristic function of a general state $\rho$ is
defined by
 \begin{align}
\chi_{\rho}(\xi)=\tr[\rho \hat{D}(\xi)]
\end{align}
in which $\xi$ is a complex parameter, $\hat{D}(\xi)=\exp(\xi \hat{a}^\dagger-\xi^*\hat{a})$ is the displacement operator \cite{Barnett3}. The characteristic function is in one-to-one correspondence with the density operator as it contains all of the information necessary to reconstruct the state. We employ this formalism throughout the paper leaving calculational details for the Appendix.

The fidelity can be increased after the comparison stage by using another mature technique known as photon subtraction \cite{Ban}, a filtration process that is implemented by allowing the output from the comparison beamsplitter to fall on a second highly transmitting ($t_2 \approx 1$) beamsplitter, with vacuum input in the second mode. The reflected mode contains a Geiger mode detector and this time the output is accepted whenever this second detector fires, an event signifying that the signal contained at least one photon. The higher the mean photon number of the input the more likely the detector is to fire, so this filters out states with lower mean photon numbers. The improvement in the output quality comes at the cost of a lower success probability of the overall scheme. If the comparison beamsplitter is 50:50 the incorrect output of the comparison stage is the vacuum, so all of the incorrect output is filtered by subtraction leaving only the amplified state $\ket {\pm g\alpha}=\ket {\pm\sqrt{2}\alpha}$, a perfect twofold gain.

\section{SCAMP for Schr\"odinger Cat States}
Motivated by the high gain and fidelity of the SCAMP for sets of individual coherent states, we investigate a modified SCAMP system for the amplification of cat states.
In the first subsection we will introduce  an updated SCAMP scheme for the amplification of cat states which only utilises Gaussian resources, beamsplitters and Geiger-mode detectors. In the second we will present the results.

\subsection{Amplification Scheme}
\begin{figure}[ht]
	\centering
	\includegraphics[trim={4cm 3.5cm 4cm 1.5cm},clip,width=10cm]{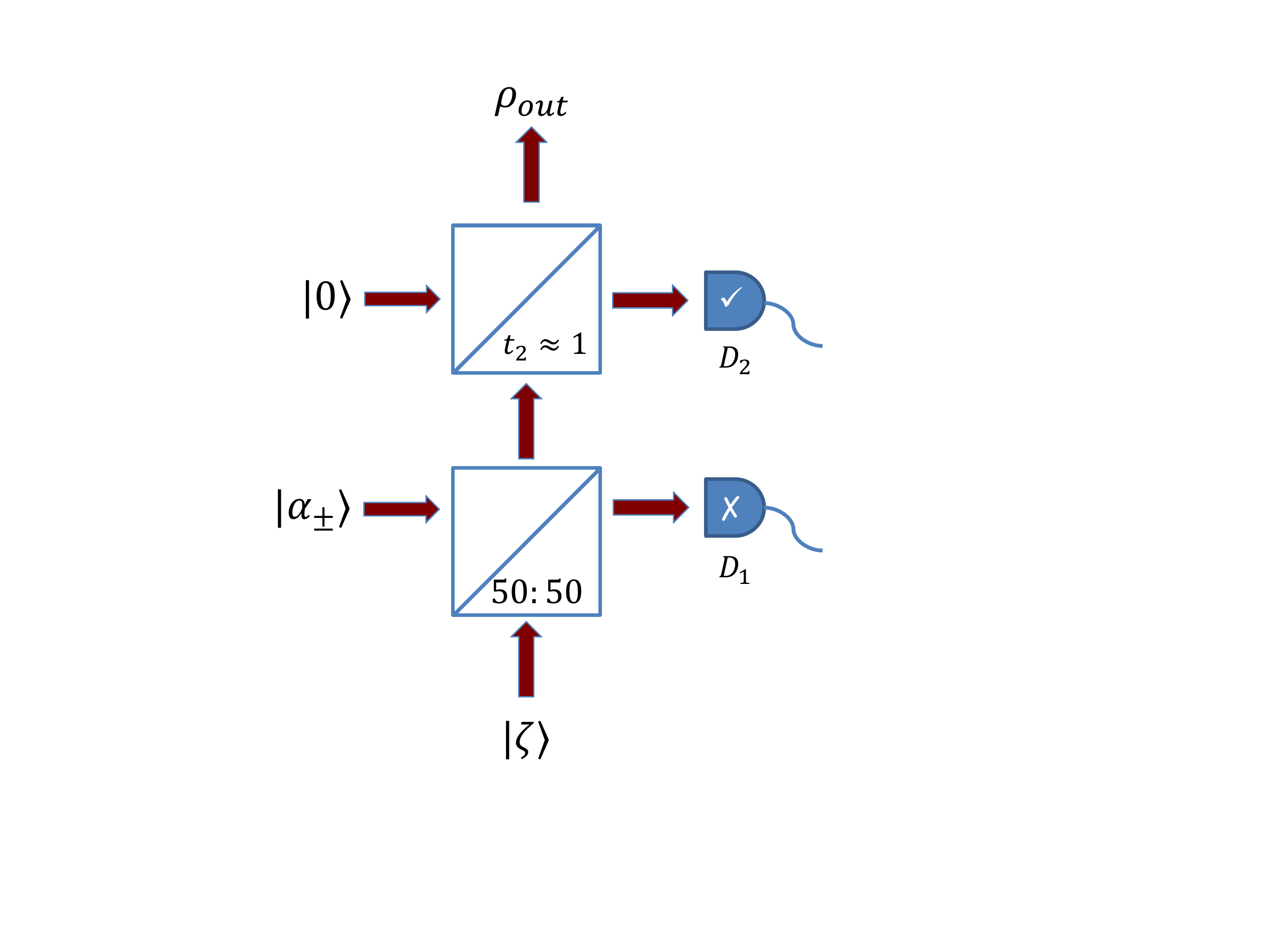}\hfill
	\caption{
	State Comparison Amplifier for Cat States comprising of a comparison stage ($50:50$ beamsplitter) and a subtraction stage.} 
	\label{catscamp}
\end{figure}

\begin{figure}[ht]
\centering
    \begin{subfigure}
        \protect{\includegraphics[width=\linewidth,height=5cm]{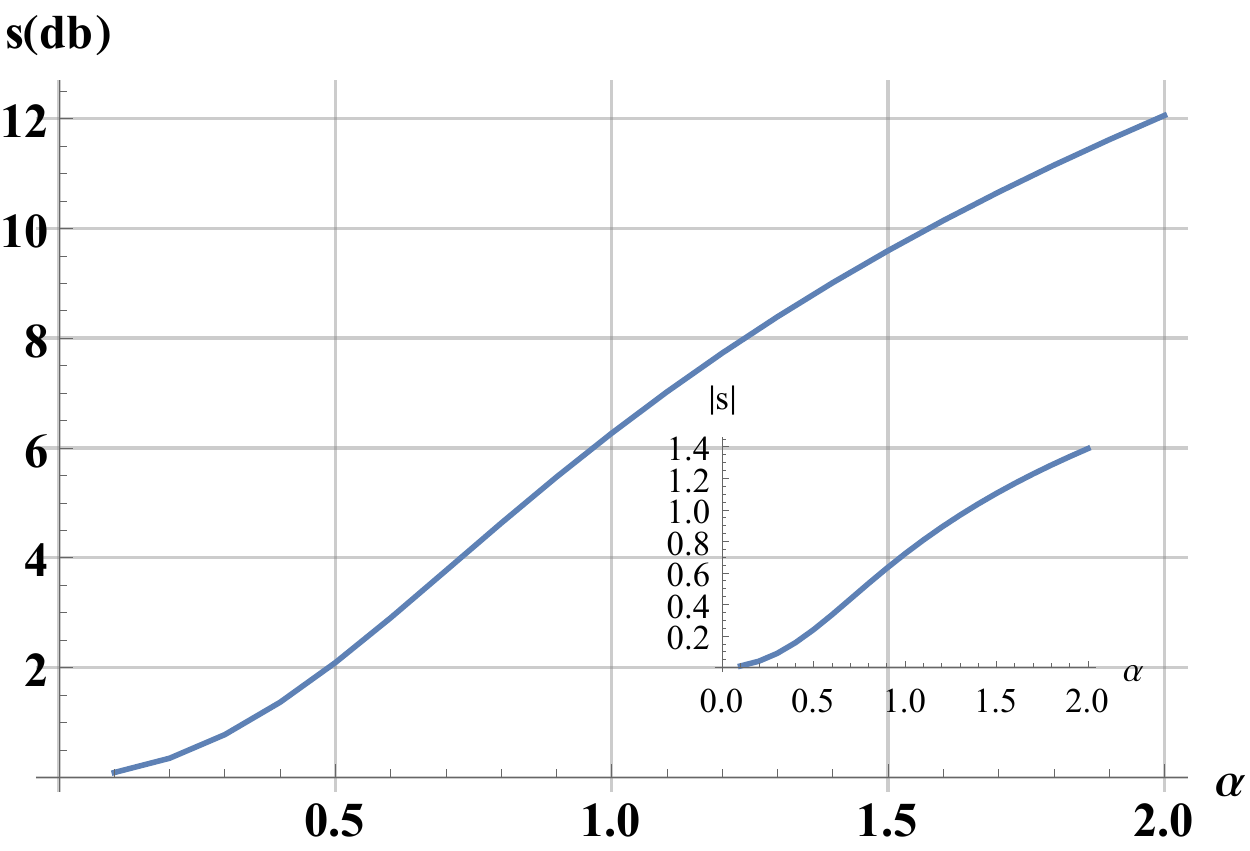}}
        \caption*{(a)}
    \end{subfigure}
    
    \begin{subfigure}
        \protect{\includegraphics[width=\linewidth,height=5cm]{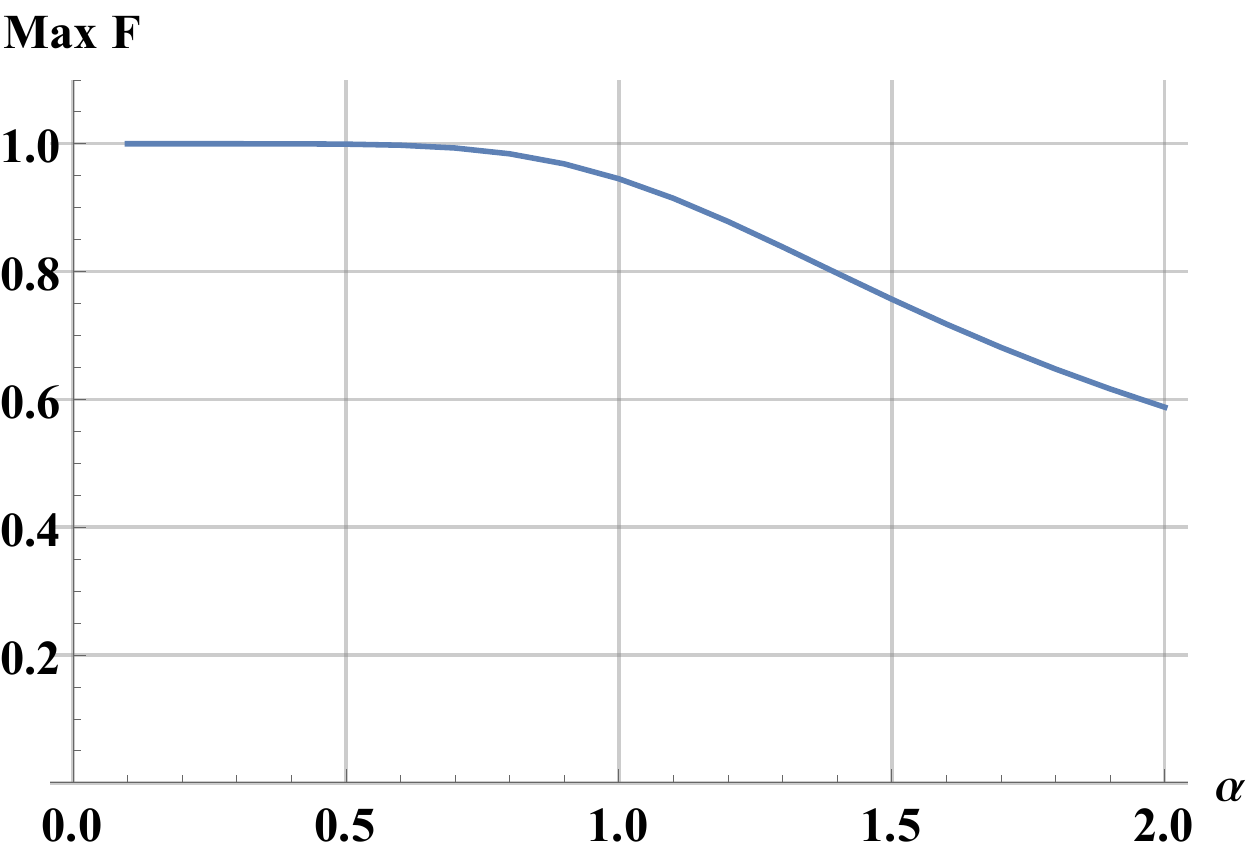}}
        \caption*{(b)}
	\end{subfigure}
		\caption{
		(a) Optimal squeezing in dB as a function of input coherent amplitude $\alpha$, ($s_{db}=-10\mbox{log}[e^{2s}]$) . 
		The inset plot shows the absolute value of the optimal squeezing in terms of the squeezing parameter. \\
		(b) Corresponding maximum fidelity as a function of input $\alpha$.}
	\label{squeezing}
\end{figure}

The amplification scheme, as one can see in Fig. \ref{catscamp} 
is an updated SCAMP in which the input state is a cat state chosen uniformly at random from the set of states $\{\ket{\alpha_+},\ket{\alpha_-}\}$ while the guess state this time is the squeezed vacuum $\ket{\zeta}$ state given by:
\begin{align}
\ket{\zeta}=\sqrt{\sech{s}}\sum_{m=0}^{\infty}\frac{\sqrt{(2m!)}}{m!}\left(-\frac{1}{2}\exp{[i\theta]}\tanh{s}\right)^m\ket{2m}
\end{align}
where $\zeta=e^{i\theta}s$, $ 0 \leq s \leq \infty $ and $0\leq \theta \leq 2\pi$.
Note that the ``guess" state is always $\ket{\zeta}$ irrespective of the input state. This particular choice is motivated by the fact that the squeezed vacuum state $\ket{\zeta}$, similarly to an even cat 
state, is a superposition of even photon numbers only, and has a high overlap with $\ket{\alpha_+}$, given by

\begin{align}
  |\langle \alpha_+|\zeta\rangle|^2= \frac{\exp{\big[-(\alpha^2+\alpha^{*2})\tanh{s}/2\big]}}{\cosh{s}\cosh{|\alpha|^2} }
\end{align}
for values of coherent amplitude up to $\alpha \approx 1$ (See Fig.\ref{squeezing} (b)).

On the other hand, the overlap of $\ket{\zeta}$ with $\ket{\alpha_-}$ is zero as the latter is a state that contains only superpositions of odd photon numbers in the photon number basis. One perhaps would expect that such a mixing of odd and even photon numbers in a beamsplitter would cause the quality of the output state to deteriorate but, as we show in the appendix, in this case such a mixing process, in conjunction with postselection in the detector arm on no counts, acts as a quantum channel that evolves the input states into squeezed states at the output.
In Fig. \ref{squeezing} we plot the squeezing magnitudes that maximise the fidelity of $\ket{\alpha_+}$ to $\ket{\zeta}$ for different values of coherent amplitude $\alpha$ and also the fidelity of $\ket{\zeta}$ with $\ket{\alpha_+}$. The squeezing values that maximise the fidelity for a given coherent amplitude admit a simple expression when we consider the coherent amplitude $\alpha$ to be real,
\begin{align}
    s=-\frac{\sinh^{-1}(2\alpha^2)}{2}.
\end{align}

The amplification protocol then proceeds as follows
\begin{enumerate}
	\item The input cat state to be amplified and the squeezed vacuum state are mixed at a $50:50$ beamsplitter and a Geiger mode detector monitors the presence or absence of light in one of the output modes.  If no event is registered, the output of this comparison stage is passed to the next stage.
	
	\item The output mode impinges on a second highly transmissive beamsplitter whose other input is the vacuum state. If the second detector registers a click the amplification is considered successful and the output is accepted. Otherwise it is discarded.
\end{enumerate}

The condition for successful amplification then can be formally given by 
\begin{equation}
P_S=P(D_1=\xmark,D_2=\checkmark)
\end{equation}

which is the joint probability of the second detector clicking and the first detector not registering a click. We show in the appendix that the comparison stage acts as a quantum channel that evolves input states to squeezed states. In the case of input cat states the output is squeezed cat states,i.e states of the form $\hat{S}(\zeta)\ket{\alpha_\pm}$ where $\hat{S}(\zeta)$ is the squeezing operator. The photon subtraction stage is needed to amplify the cat states. The details of the computation have been left for the appendix.

Fig. \ref{squeezing} shows that the amount of squeezing required to optimise the fidelity between the output and an ideal cat state for a wide range of input cat state sizes is moderate. For example, if $|\alpha|^2 =1$ the amount of squeezing required is 6 dB - a factor of 4. Even for $|\alpha|^2 = 4$  a squeezing of 12 db ($|s|\leq 1.4$) is required, which has been experimentally generated in a doubly resonant, type I optical parametric amplifier (OPA) operated below threshold \cite{Vahlbruch}. 

\subsection{Results}

Here we present the results of fidelity, gain and probability of success to benchmark the amplifier. We provide the derivation of these results in the Appendix. We have assumed that the coherent amplitudes of both the output states and the nominal states are real, without loss of generality.

The performance of the amplifier is benchmarked using the fidelity $\mbox{F}$ between the output state and an ideal amplified cat state $\{\ket{g\alpha_+},\ket{g\alpha_-}\}$. We note that photon subtraction changes the parity of the cat state \cite{Dakna}. 
 To see why that is the case, consider applying the annihilation operator, $\hat{a}$, to the even cat state, $\ket{\alpha_+}\sim \ket{+\alpha}+\ket{-\alpha}$; coherent states are eigenstates of $\hat{a}$ satisfying $\ket{\alpha}=\alpha\ket{\alpha}$ and thus we get

\begin{align}
\hat{a}\ket{\alpha_+}\sim \hat{a}\ket{+\alpha}+\hat{a}\ket{-\alpha}=\alpha (\ket{+\alpha}-\ket{-\alpha})
\end{align}
The right hand side of the equation is an (un-normalised) odd cat state $\ket{\alpha_-}\sim \ket{+\alpha}-\ket{-\alpha}$.
So for example if we start with an even cat state, after photon subtraction we obtain an odd cat state and this subtraction occurs in the limit $t_2 \approx 1$. The application of $\hat{a}$ to states of even photon number produces states of odd photon number and vice versa.

\begin{figure}[ht]
    \begin{subfigure}
		\protect{\includegraphics[width=\linewidth,height=5cm]{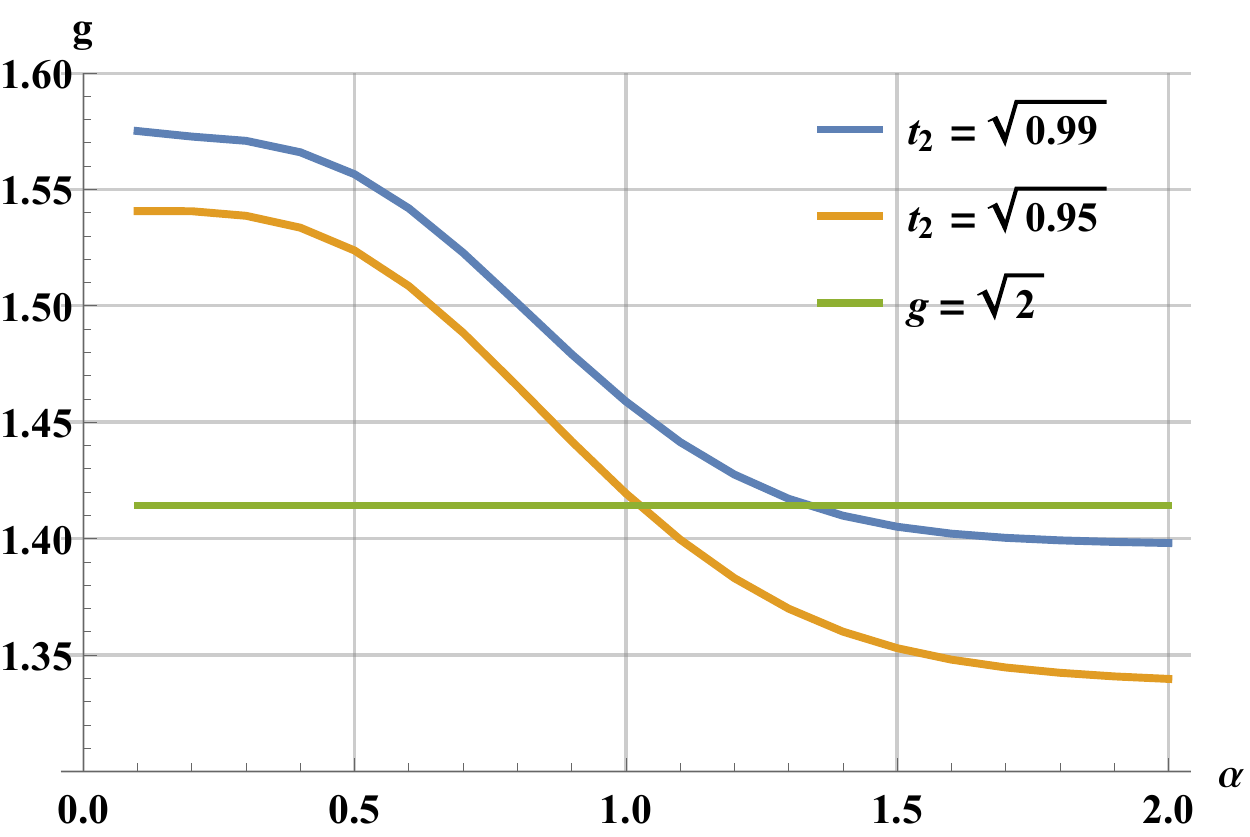}}
		\caption*{(a)}
	\end{subfigure}
		
	\begin{subfigure}
		\protect{\includegraphics[width=\linewidth,height=5cm]{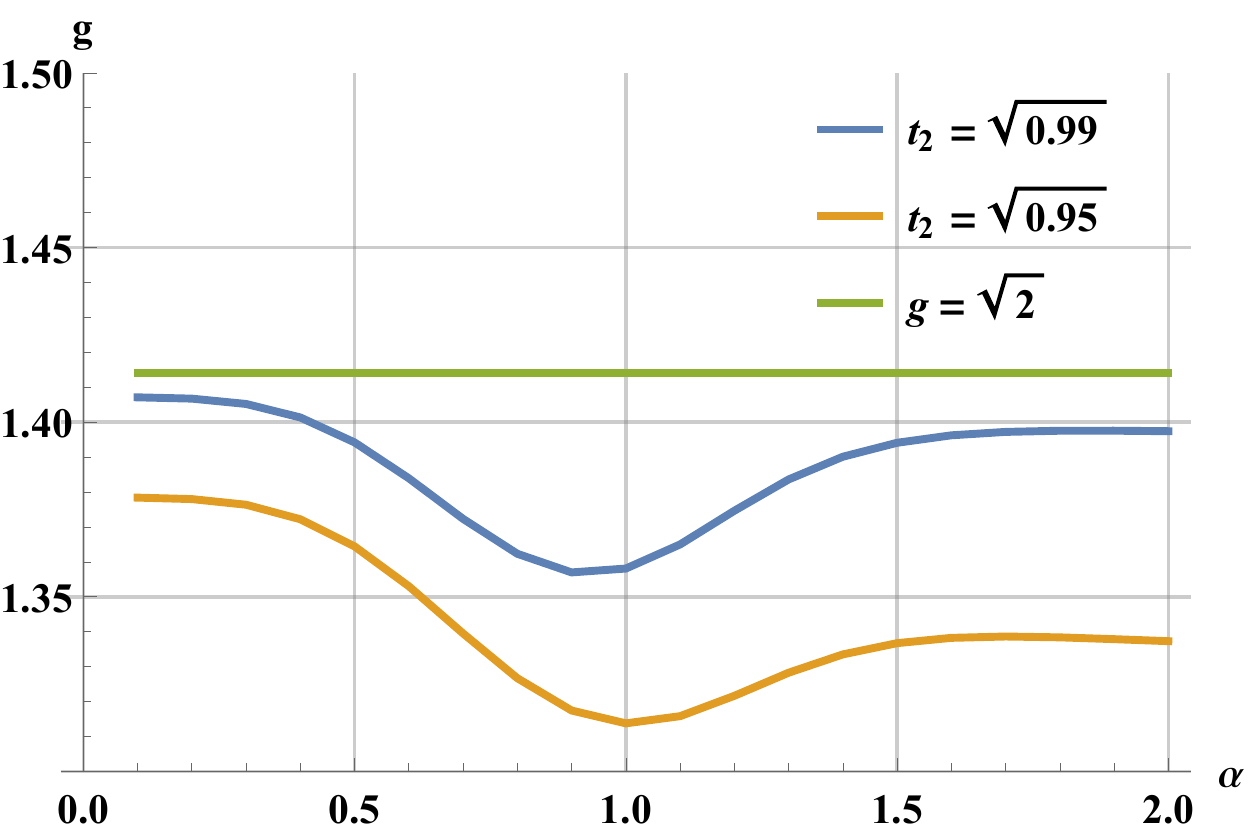}}
	    \caption*{(b)}
	\end{subfigure}
	\caption{(a) Optimal gain g as a function of the size $\alpha $ of the input even cat state. (b) Same plot for an odd cat state. The green line indicates $\sqrt{2}$ gain.}
	\label{gain}
\end{figure}

\begin{figure}[ht]
    \begin{subfigure}
        \protect{\includegraphics[width=\linewidth,height=5cm]{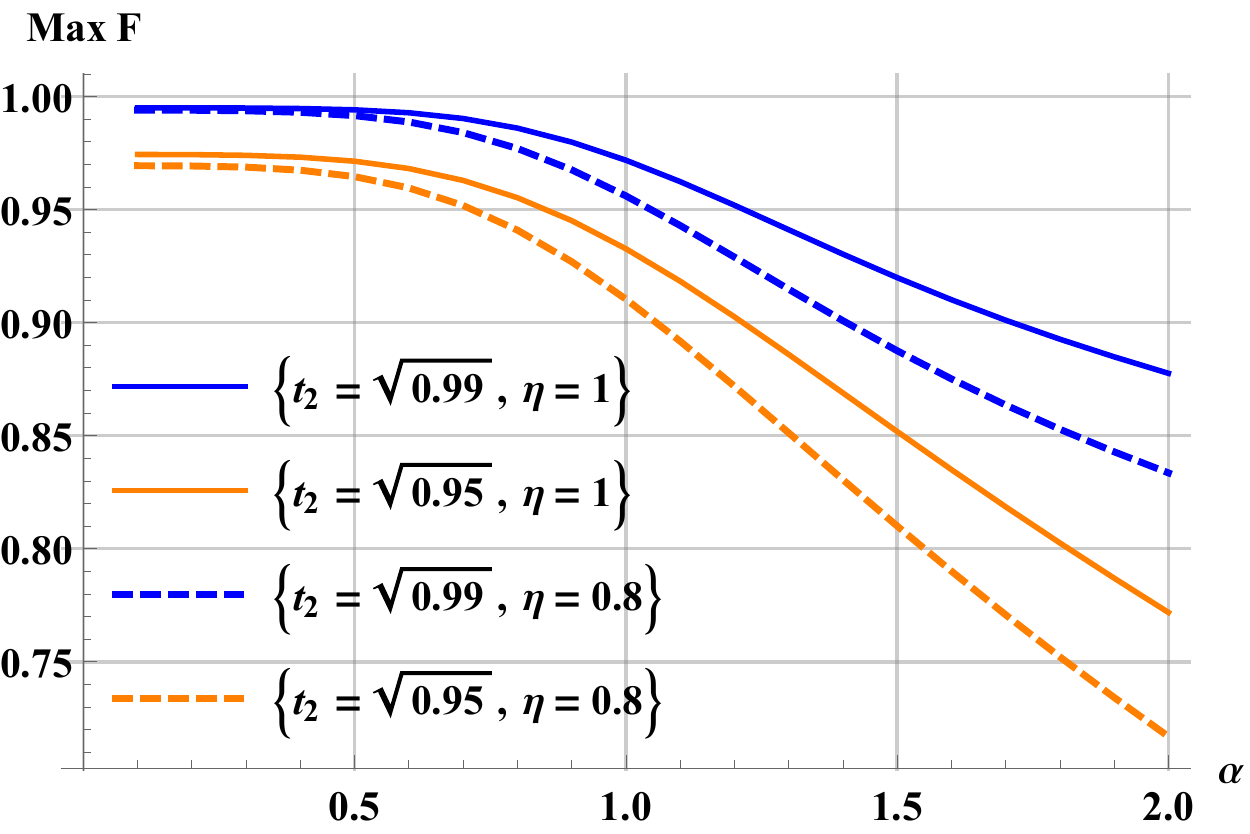}}
		\caption*{(a)}
    \end{subfigure}
    
    \begin{subfigure}
	    	\protect{\includegraphics[width=\linewidth,height=5cm]{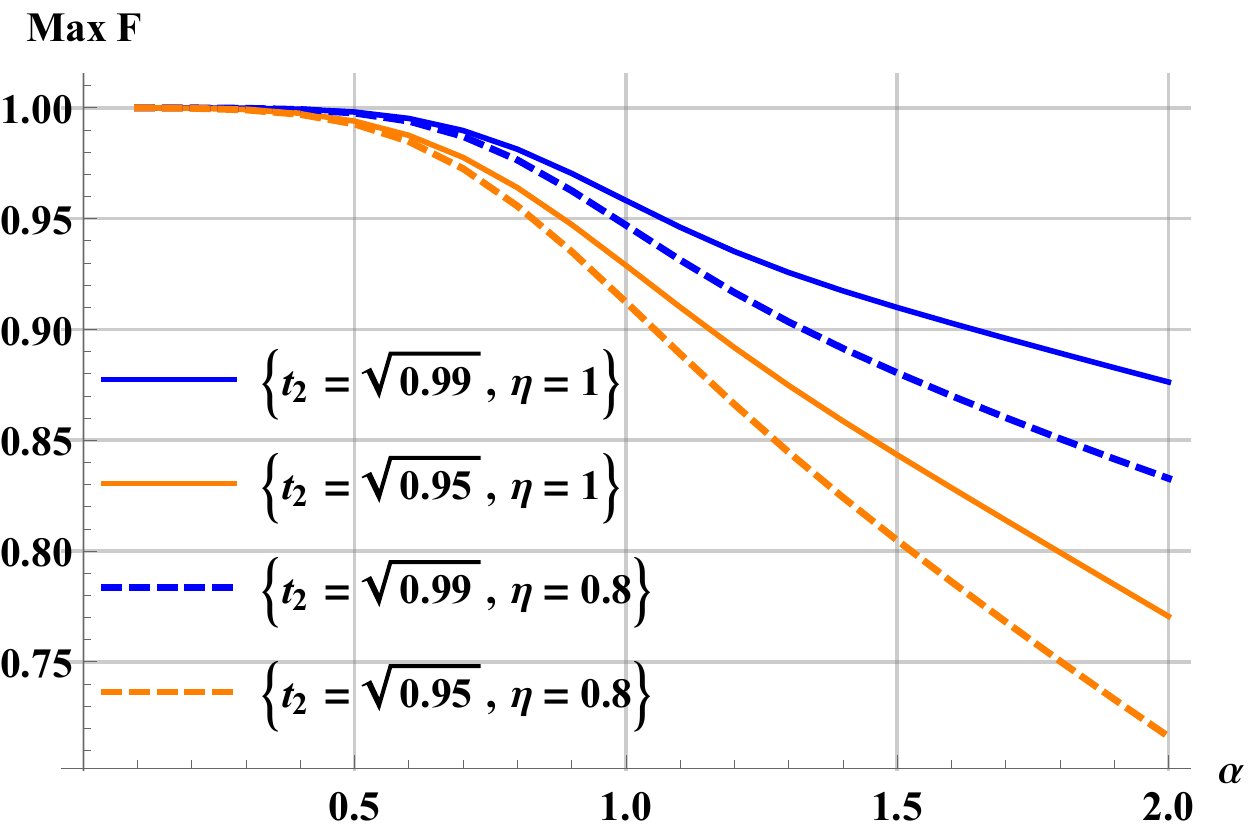}}
		\caption*{(b)}
	
	\end{subfigure}
	
	\caption{(a) Optimal fidelity as a function of the size $\alpha $ of the input even cat state.(b) Same plot for an odd cat state.}
	\label{fidelity}
\end{figure}

\begin{figure}[h!]
    \begin{subfigure}
	\protect{\includegraphics[width=\linewidth,height=5cm]{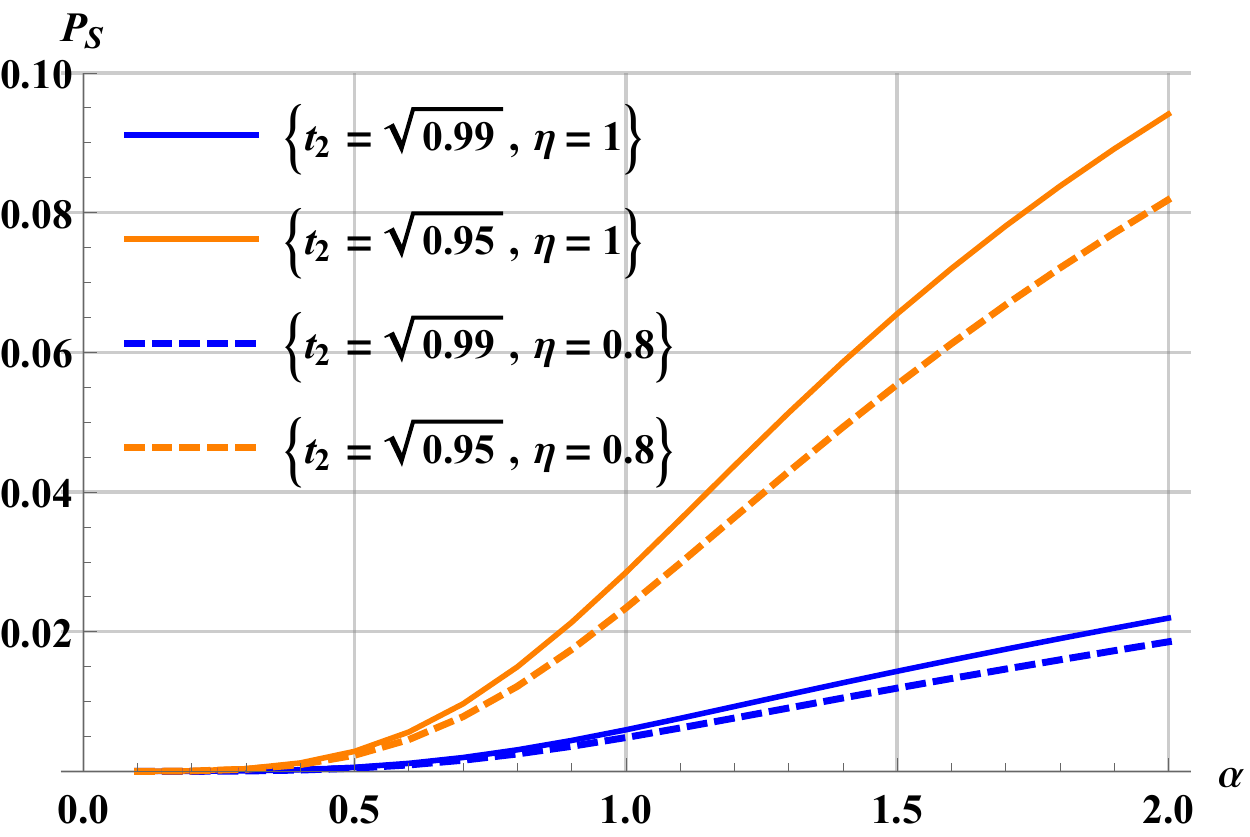}}
	\caption*{(a)}
	\end{subfigure}
	
	\begin{subfigure}
		\protect{\includegraphics[width=\linewidth,height=5cm]{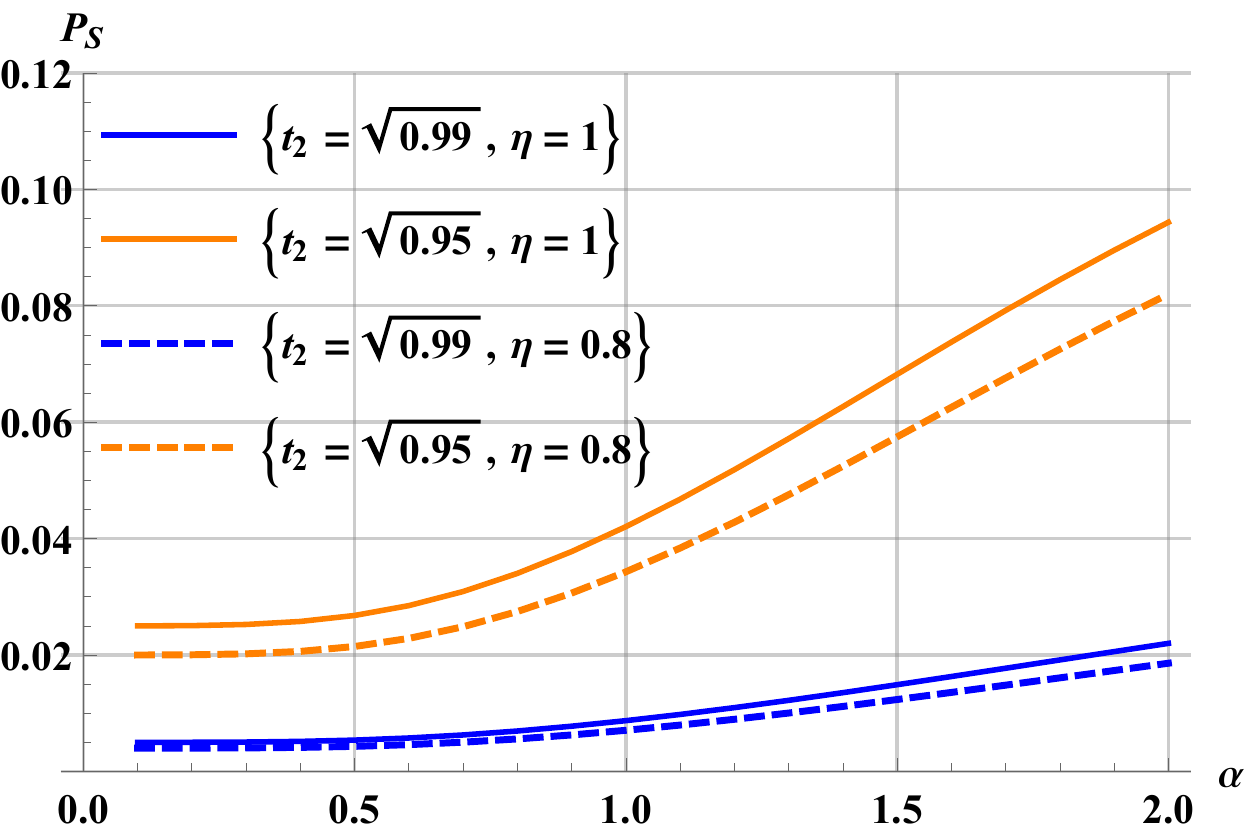}}
	\caption*{(b)}
	\end{subfigure}
	\caption{(a) Corresponding success probability as a function of the size $\alpha $ of the input even cat state. (b) Same plot for an odd cat state.}
	\label{probability}
\end{figure}

In Fig. \ref{gain}-\ref{probability} we summarise results for the gain, maximum fidelity and probability of success for the output state after the amplification process. In the following we have assumed that the dark counts in the detection process are negligible (they can be made so in pulsed systems by time filtering around the pulse centre) and we have considered two scenarios where the quantum efficiencies for the two detectors are both ideal ($\eta_1=\eta_2=1$) or both detectors are $80\%$ efficient($\eta_1=\eta_2=0.8$). 

 Fig. \ref{gain} is a plot of the amplitude gain for different values of input cat state size. The curves shown are for ideal detector efficiency as quantum efficiency does not affect the gain of the amplifier. The gain does depend on the transmission coefficient of the second beamsplitter as this quantifies the quality of the photon subtraction. A higher transmission coefficient leads to a higher gain as the photon subtraction stage is more faithfully approximated as $t_2 \approx 1$; this comes at the expense of the success probability as in this limit there is hardly any light reflected to trigger the second detector. One can see that there is an approximately twofold intensity gain for both input cat states and for all values of coherent amplitude considered here. This may seem counter-intuitive for the case where the input is an odd cat state as the state has vanishing overlap with the squeezed vacuum state, but we should remark that the gain in this case, contrary to the original proposal of SCAMP, is not provided by the nullification of the signal in one output mode after the comparison stage but is a combination of effects from both the comparison and the photon subtraction stages (see next section for details). 

The fidelity (Fig. \ref{fidelity}) depends both on the transmission coefficient and the quantum efficiency of the detectors. The higher both quantities are the better the fidelity of the output with the corresponding ideal output cat state. The fidelity also depends on the input cat state size and it is near unity for small values of the input coherent amplitude $\alpha$. It is greater than 80 \% for input $\alpha$ up to $\alpha\approx 1.5$ which corresponds to an output cat state of amplitude $\beta=1.95$ if the transmission coefficient is $t_2=\sqrt{0.95}$ and the detector efficiencies are  $\eta=0.8$. 

Not surprisingly, the lower transmission coefficient 
of the second beamsplitter (higher reflection coefficient) increases the probability of success as the probability of the second detector clicking depends on the amplitude of the reflected light that impinges the detector. For the same reason the success probability increases with increasing cat input amplitude. Overall, the amplifier works with a high fidelity and approximately two-fold intensity gain for a range of input cat state sizes up to $\alpha \approx 1.5$ corresponding to an output cat state of size $\beta=1.95$. Considering, the most realistic implementation where ($t_2=\sqrt{0.95},\eta=0.8$) and keeping in mind the practical requirement that input cat states of size $\alpha \geq 1.2$ are necessary, the Parity Swap SCAMP for Cat States can provide an output state that has a fidelity of $87\%$ with an ideal cat state of size $\beta=1.5$, requiring an input state of size $\alpha=1.1$ with an overall success probability of $ \approx 3\%$ without prior knowledge of the parity of the input state. In other words, the amplifier transforms an input cat state to an amplified output cat state of opposite parity of approximately double the mean photon number of the input and works symmetrically on both even and odd cat states.

\begin{figure}[h]
    \begin{subfigure}
        \protect{\includegraphics[width=\linewidth,height=5cm]{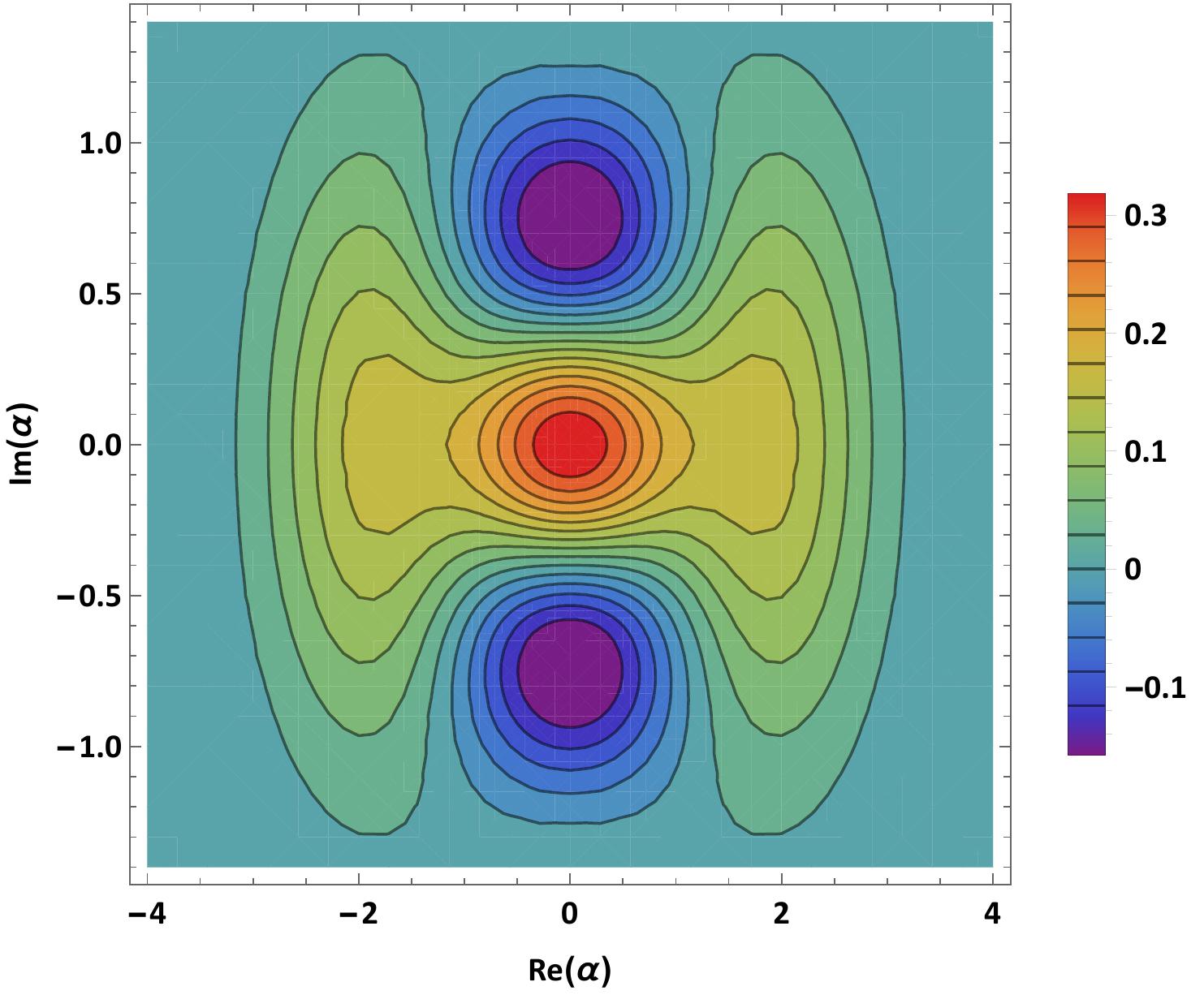}}
		\caption*{(a)}
    \end{subfigure}
    
    \begin{subfigure}
	    	\protect{\includegraphics[width=\linewidth,height=5cm]{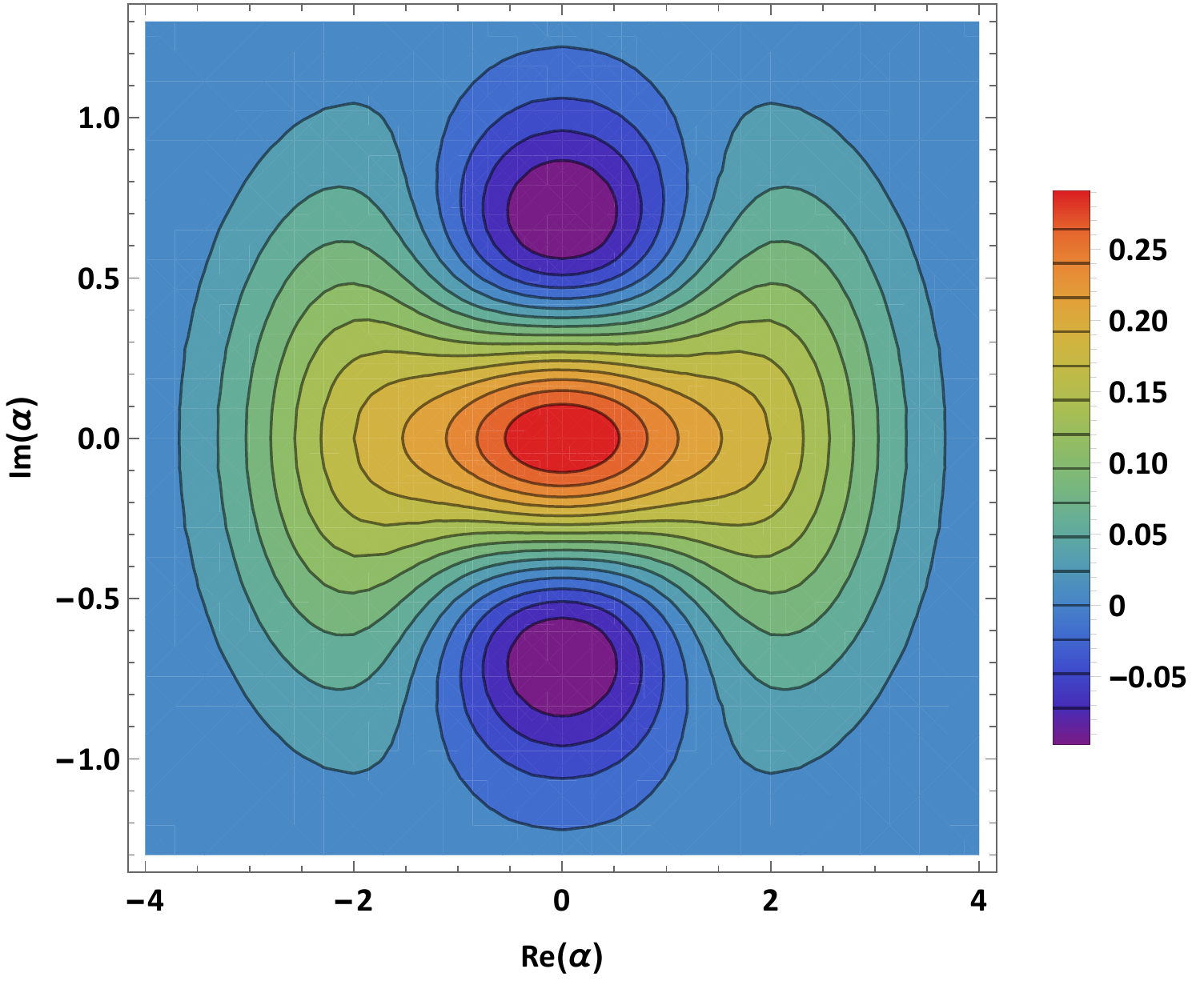}}
		\caption*{(b)}
	
	\end{subfigure}
	
	\caption{(a) Contour plot of the Wigner function of an even cat state of size $\alpha=1.31$ (b) Same plot for the output state of the swap parity cat scamp given an odd cat state input of size $\alpha=1$.}
	\label{contourOdd}
\end{figure}

\begin{figure}[h]
    \begin{subfigure}
        \protect{\includegraphics[width=\linewidth,height=5cm]{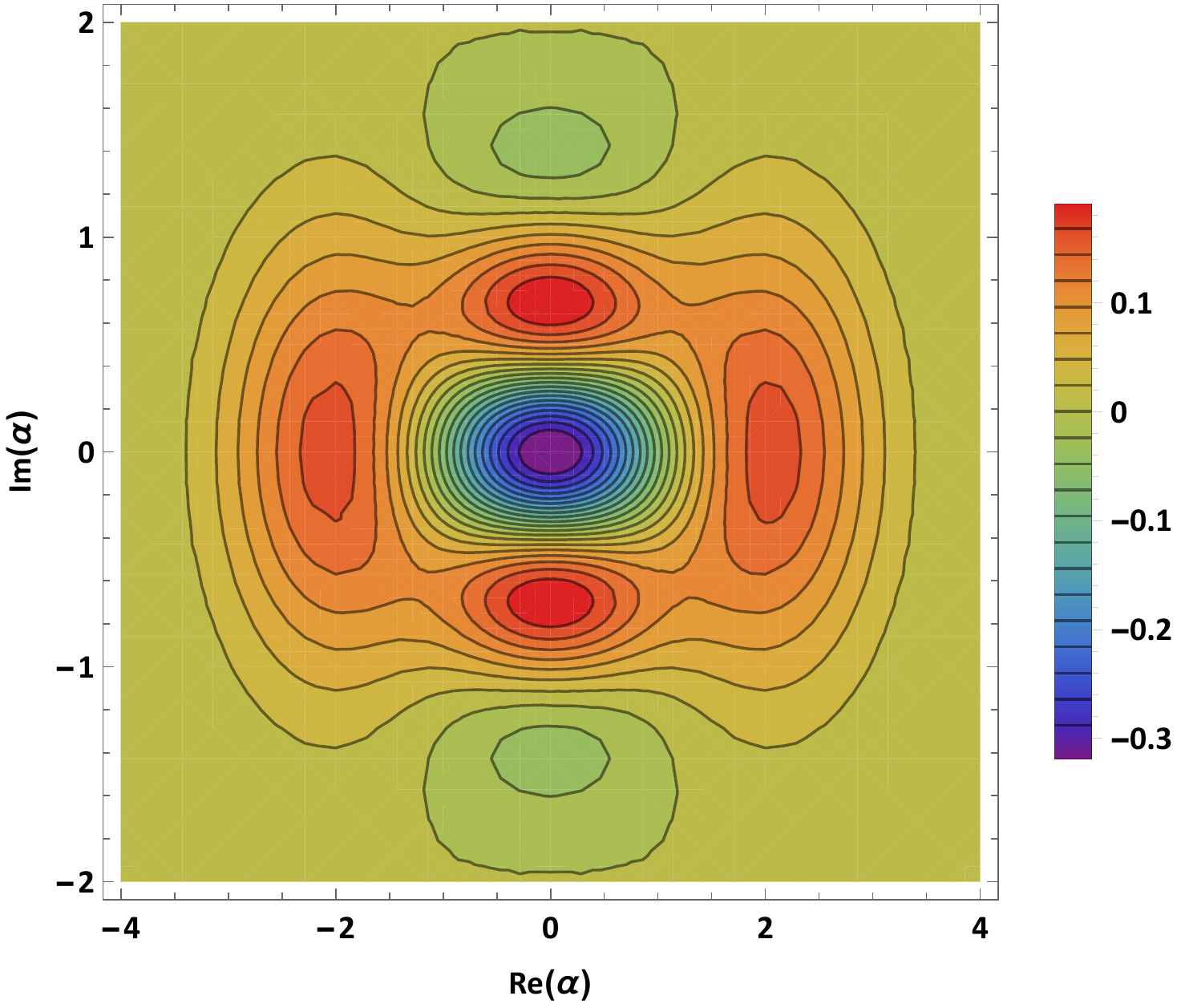}}
		\caption*{(a)}
    \end{subfigure}
    
    \begin{subfigure}
	    	\protect{\includegraphics[width=\linewidth,height=5cm]{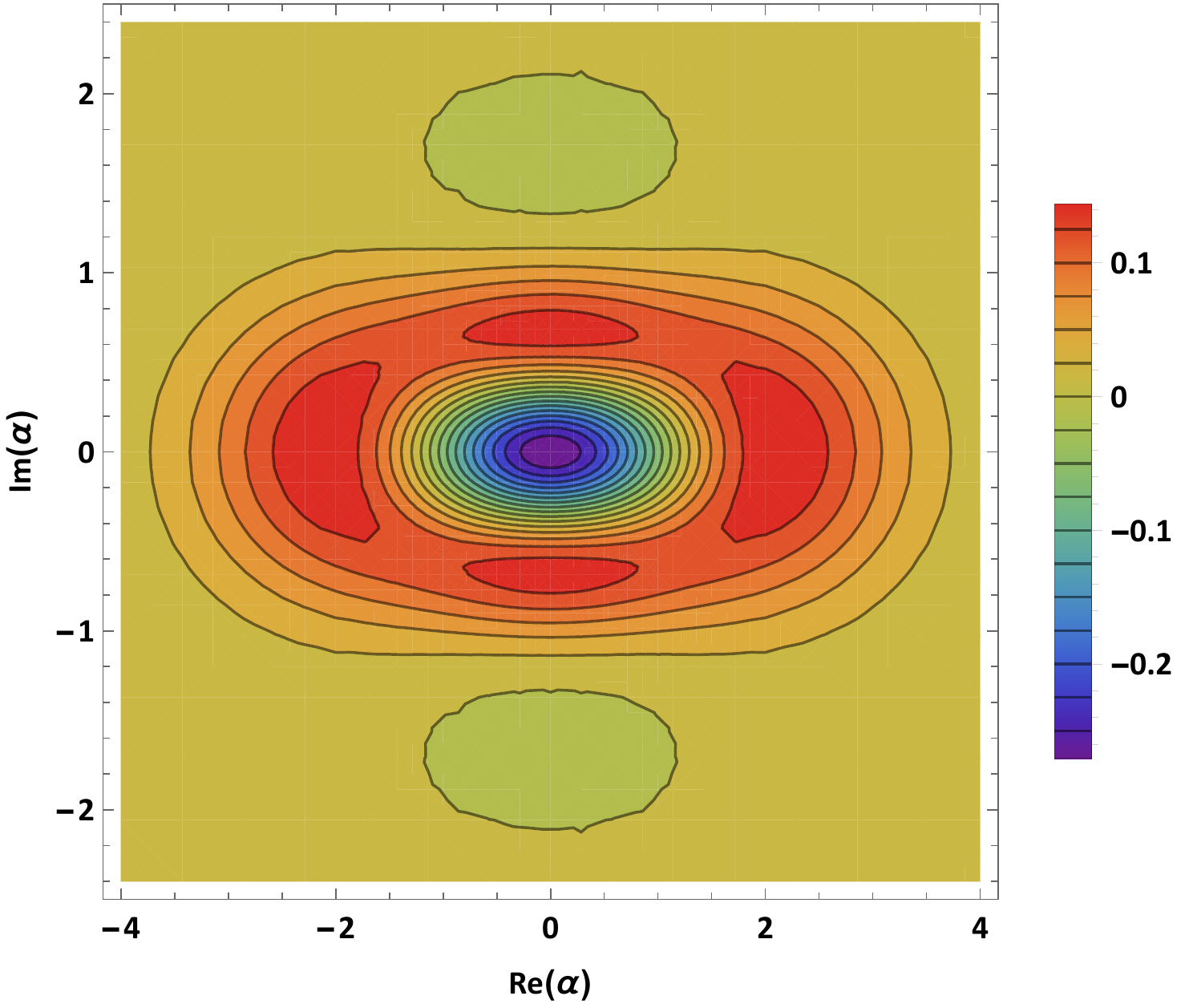}}
		\caption*{(b)}
	
	\end{subfigure}
	
	\caption{(a) Contour plot of the Wigner function of an odd cat state of size $\alpha=1.42$ (b) Same plot for the output state of the swap parity cat scamp given an even cat state input of size $\alpha=1$.}
	\label{contourEven}
\end{figure}

The fidelity is one indication of the output quality but it does not contain any information about certain features of the quantum states, such as those contained in the form of the Wigner function, of which one example might be negativity - a signature of nonclassicality. It is therefore instructive to plot the contour plots of the output states after the amplification process and of the equivalent ideal amplified cat states. In Figs. \ref{contourOdd}, \ref{contourEven} we plot the Wigner functions of the output amplified state when the inputs are an even and an odd cat state of size $\alpha=1$ and alongside the Wigner functions of the ideal amplified cat states. We choose the experimentally feasible parameters of $t_2=\sqrt{0.95}, \eta=0.8$ where the fidelity between the output and the ideal cat state in both cases is $91\%$.

The main features of the cat states in phase space, such as the distinguishability of the constituent coherent states and the interference features of the superposition are faithfully preserved. This provides a more qualitative assessment of the output states which compliments the quantitative benchmark of the fidelity metric. When the input state to the device is an odd cat state, the negativity of its Wigner function, compared to the negativity of the Wigner function of the ideal amplified even cat state, is lower (by approximately a factor of 2), but for an even input cat state it is comparable. 
\section{Theoretical Considerations}
To understand how SCAMP can be used to amplify Schr\"odinger cat states we need to consider how photon subtraction affects a squeezed cat state. A squeezed cat state can be generated by applying the squeezing operator $\hat{S}(\zeta)$ to a cat state $\ket{\alpha_\pm}$. Photon subtraction then leads to

\begin{align}
    \nn \hat{a}\hat{S}(\zeta)\ket{\alpha_\pm}=&\hat{S}(\zeta)\hat{S}^\dagger (\zeta)\hat{a}\hat{S}(\zeta)\ket{\alpha_\pm}\\
    =&\hat{S}(\zeta)(\cosh{s} \hspace{0.1cm} \hat{a}\ket{\alpha_\pm}-\sinh{s}\hspace{0.1cm} \hat{a}^\dagger\ket{\alpha_\pm}),
\end{align}
which is a squeezed superposition of a photon subtracted and a photon added cat state. While photon subtraction simply swaps the parity of the cat state, the photon addition swaps the parity and increases the amplitude of the state \cite{Barnett}. One may wonder whether photon addition alone would work but we should note that a photon added even cat state is a vacuum removed state. Therefore, the fidelity between an ideal even cat state and the photon added cat state would be lower than the fidelity between the output state of the cat SCAMP, which produces a state that contains a vacuum component.  The combined effect of photon addition and photon subtraction, rather than state comparison, is the main gain mechanism of SCAMP for cat states. In principle we could undo the squeezing by a local unitary operation and we would only be left with a superposition of a photon subtracted and a photon added cat state, but it may be challenging to implement experimentally.

The overlap between an ideal cat state and the output of the amplification scheme is 
\begin{align}
    \bra{\beta_\mp}\hat{a}\hat{S}(\zeta)\ket{\alpha_\pm}=\nonumber & N_\pm^{-\frac{1}{2}} \Big[  2\sqrt{\sech{s}}\\ \nonumber
    & e^{-|\beta-\gamma|^2/2+(-\beta \gamma^*+\beta^* \gamma)/2-\tanh{s}(\beta^*-\gamma^*)^2/2}\\ \nonumber
    &(\gamma-(\beta^*-\gamma^*)\tanh{s})\\
    \nonumber
    & +2\sqrt{\sech{s}}\\ \nonumber
    & e^{-|\beta+\gamma|^2/2+(\beta \gamma^* -\beta^* \gamma)/2-\tanh{s}(\beta^* + \gamma^*)^2/2}\\
    &(\pm \gamma\pm (\beta^*+\gamma^*)\tanh{s}) 
    \Big]
    \end{align}
    where $\gamma=\alpha \cosh{s}-\alpha^* \sinh{s}$ and $N_\pm$ is given by
    \begin{align}
    \nonumber
    N_\pm=&(2\mp 2e^{-2|\beta|^2})\\
    \nonumber
    &\big\{2|\alpha|^2(1\mp e^{-2|\alpha|^2})(\cosh{s}^2+\sinh{s}^2)\\ \nonumber
    &+(2\pm 2e^{-2|\alpha|^2})\\
    &(\sinh{s}^2-2\cosh{s}\sinh{s}Re[\alpha]^2)\big\}
    \end{align}
The fidelity is then simply $\mbox{F}=| \bra{\beta_\mp}\hat{a}\hat{S}(\zeta)\ket{\alpha_\pm}|^2$.

The state after the comparison stage and upon postselection on zero clicks, assuming a perfect detector, is a squeezed cat state of a scaled squeezing parameter and a scaled coherent amplitude, both smaller than the equivalent quantities of the input states. The squeezing parameter $s^\prime$ is given by

\begin{align}
s^\prime=\ln{\sqrt{\frac{\cosh{s}+(1-r_{1}^2)\sinh{s}}{\cosh{s}-(1-r_{1}^2)\sinh{s}}}}
\end{align}
 where s is the squeezing parameter of the input squeezed vacuum state and $r_1$ the reflection coefficient of the beamsplitter of the comparison stage.
 
 The scaled coherent amplitude, $\alpha^\prime$ of the state after the comparison stage is given by 
  \begin{align}
      \alpha^\prime=\frac{r_1 \alpha \cosh{s}}{\sqrt{(\cosh{s})^2-(1-r_{1}^2)^2(\sinh{s})^2}}
  \end{align}
 Then we can plot the ratio of the ideal cat size $\beta$, which maximises the fidelity to the photon subtracted squeezed cat state, to the input cat size $\alpha$ as a function of the input cat size $\alpha$.
 
 \begin{figure}[H]
   \includegraphics[width=\linewidth,height=5cm]{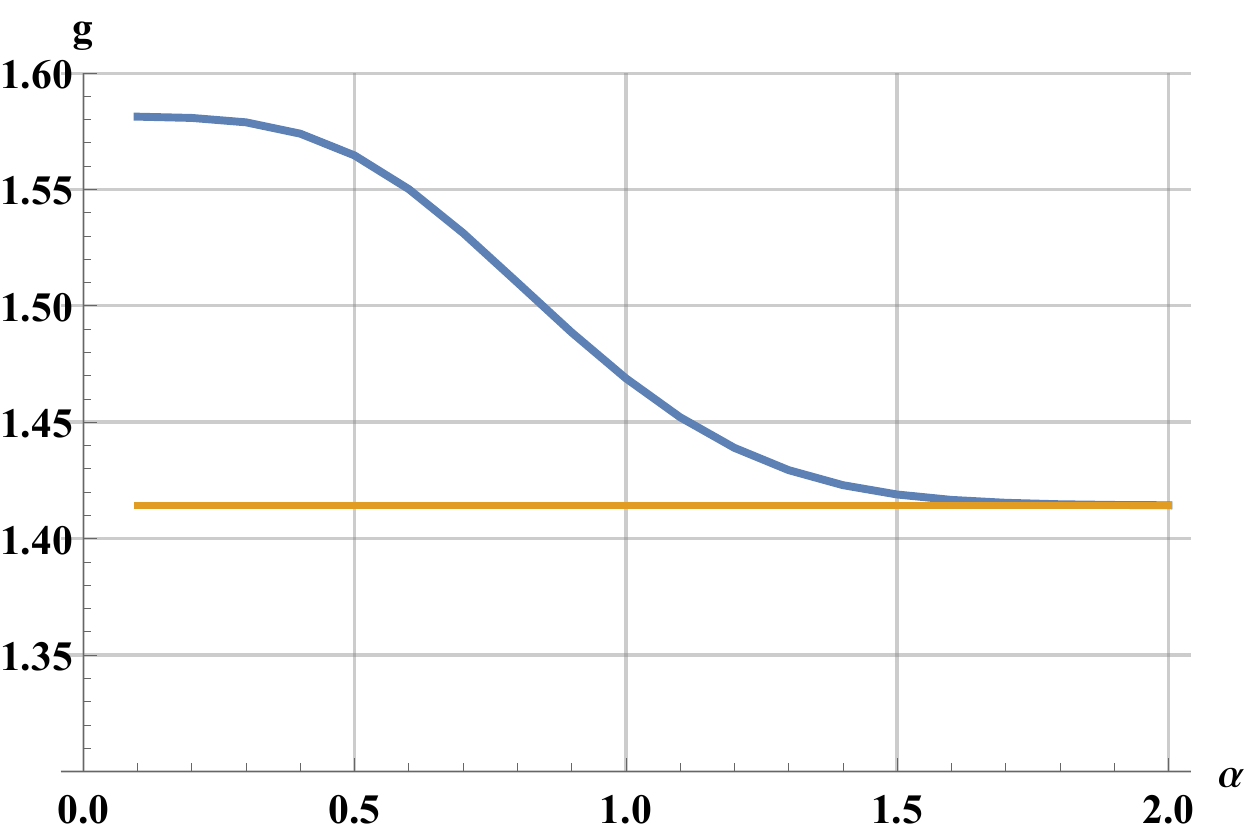}
	\caption{ Ideal gain given by subtracting a photon from a squeezed even cat state as a function of input cat size.}
	\label{IdealGain}
\end{figure}

One can indeed verify that Fig. \ref{gain} (a) and Fig. \ref{IdealGain} differ marginally due to the fact that photon subtraction can only be implemented approximately and is limited by the transmission coefficient of the beamsplitter used to perform the photon subtraction (which is always less than unity, $t_2=\sqrt{0.99}$ in this case).

In this section we have described the theoretical mechanism for the gain of our device. Physically it is based on mean photon number matching between a resource state (squeezed vacuum) and the input cat state after photon subtraction conditioning. If the cat state  has a significantly higher photon number than the resource state the output will effectively be a photon subtracted cat state. If the resource state has a significantly higher photon number than the cat state the output will be a photon subtracted resource state. Only if they are comparable in mean photon number does the significant state overlap between the two give an amplified cat. The approximate photon number matching of the two states is also the reason for the approximate twofold gain.

\section{Other Schemes for Amplification of Schr\"odinger Cats}

Figure 9 suggests that we could amplify a Schr\"odinger cat state reasonably well by first squeezing the cat state and then subtracting a photon. The process amounts to a slightly simpler theoretical scheme than ours. However, squeezing low amplitude non-vacuum states in a controlled fashion is more difficult experimentally, particularly in the pulsed domain. Optical nonlinearities are small and require high pump fields with a good mode overlap with the signal. For these reasons we believe that such a direct-drive scheme is a little less experimentally practical than ours.

The state of the art experimental scheme of \cite{Synchev} which utilises the cat ``breeding" method produces an amplified cat state with amplitude $\alpha=1.85$ while consuming two cat states of amplitude $\alpha=1.15$ as resources. The probability of success is $P_S=0.2$ and the fidelity of the produced state with an ideal cat state is $\mbox{F}=0.77$. Seemingly, SCAMP is on par with cat ``breeding" schemes in terms of fidelity and gain while it offers a lower probability of success. One has though to take into account the fact that cat ``breeding" schemes consume two nongaussian resource states, as opposed to just one for SCAMP, that have a low success probability of production currently and therefore offset this advantage. If push-button quantum state generation becomes a reality then the intrinsic advantages of cat ``breeding" may be more easily realised. 
 
\section{Conclusions}
In summary, we have presented a theoretical scheme for the amplification of input Schr\"odinger cat states based on a nondeterministic amplifier (SCAMP) comprised of the two mature techniques of state comparison and photon subtraction. The device both amplifies and swaps the parity of a state chosen from the set of odd and even Schrodinger cat states. Hence we call it the parity-swap cat state comparison amplifier.

The implementation of SCAMP for Schr\"odinger cat states would require only a Gaussian resource state (squeezed vacuum), linear optical components and Geiger mode detectors, making it experimentally feasible. The resource state is not randomly chosen, as it is in the standard SCAMP for coherent state amplification. This, coupled with the fact that it is Gaussian gives a significant advantage over other schemes. The SCAMP itself provides the photon subtraction required to render the output nongaussian. The parity-swap cat SCAMP can work almost symmetrically for both even and odd cat states without prior knowledge of the input state and offers high fidelity and reasonably high gain for the range of input cat state sizes of interest.

We have characterised the performance of the amplifier via gain, fidelity and success probability. The intensity gain is shown to be approximately twofold for both the even and odd cat states for a wide range of mean input photon numbers, reflecting the fact that the gain of a standard SCAMP is approximately two for a 50/50 comparison beamsplitter. The fidelity of the output with an ideal amplified cat state $\ket{g\alpha_\pm}$ depends on the transmission coefficient, $t_2$, of the beamsplitter  used to perform the photon subtraction stage. The fidelity remains high ($F \geq 81\%$)) for values of input $\alpha$ up to $\alpha \approx1.5$ in realistic parameter ranges. As an example, one can amplify an input cat state of size $\alpha=1.1$ which, after the amplification, has a fidelity of $87\%$ with a cat state of size $\beta=1.5$ and the overall success probability is approximately $3\%$. The probability of success is comparable with schemes that consume non-Gaussian resources. However, when one takes into account the overall experimental probability of producing the non-Gaussian resources, which is significantly lower compared to Gaussian ones, the scheme presented here has a clear advantage. 

There are a couple of limitations to the usefulness of the scheme. The low success probability is a common problem in the quantum regime for all schemes that rely on photon subtraction. The original SCAMP for coherent states has a similar success probability in this regime. This is compensated for by running the experiment at a high rate and the same technique would work here, although the limiting speed may be the cat-state production rate for the input states.  Secondly, the resource state, the squeezed vacuum, contains only even photon numbers and so the parity-swap cat SCAMP can only work for cat states of even or odd photon number. States such as 
\bea
\ket{\alpha_{\pm i}} = N \left(\ket{\alpha} \pm i \ket{-\alpha} \right)
\eea
do not satisfy this criterion and cannot be amplified in this fashion. However, such states have never been produced in the laboratory and so are of limited interest.

\section{Acknowlegments}
The authors would like to acknowledge financial support from the UK National Quantum Technology Programme via EPSRC and the Quantum Technology Hub in Quantum Communications (Grant EP/M013472/1). GT would also like to acknowledge EPSRC for a partial studentship from the Doctoral Training Grant to the University of Strathclyde.

\section*{Appendix}
In this appendix section we will use the formalism of characteristic functions to sketch the method used to derive all of the required results. We will see how the states and operators can be represented under the formalism and show how one can compute the benchmark metrics considered in the main body of the text.

\subsection*{Characteristic Functions}

A useful feature of the characteristic function formalism is that the trace of operators can be evaluated as an integral in phase space. More formally, the trace rule for two operators $\hat{O}_1$ and $\hat{O}_2$ is given by \cite{Ferraro}
 
\begin{equation}
\mbox{Tr}[\hat{O}_1\hat{O}_2]=\int_{\mathbb{C}} \frac{d^{2}\xi}{\pi} \chi_{\hat{O}_1}(\xi)\chi_{\hat{O}_2}(-\xi)
\end{equation}
and this formula gives us the fidelity \mbox{F}.
States whose characteristic function is a Gaussian are known as Gaussian states and one such example we have already seen; namely the coherent state $\ket{\alpha}$ whose characteristic function is
\begin{align}
\chi_\alpha(\xi)=
\exp(\xi \alpha^*-\xi^* \alpha-|\xi|^2/2)
\end{align}
and the squeezed vacuum state whose characteristic function is 
\begin{align}
\chi_{\zeta}(\xi)=&\exp\Big(-\xi^2_r\exp[2s]/2 +\xi^2_i\exp[-2s]\Big)
\end{align}
where $\xi_r$ and $\xi_i$ are the real and imaginary parts of the complex variable $\xi$.

The cat states on the other hand are nongaussian states as their characteristic functions are given by sums of Gaussians:
\begin{align}
\chi_{\alpha_+}(\xi)=&N_{+}^2\Big [\exp\left(\xi  \alpha ^*-\alpha  \xi ^*- |\xi|^2/2  \right)\nonumber\\&+\exp\left(-\xi  \alpha ^*+\alpha  \xi ^*- |\xi|^2/2\right)\nonumber\\
&+\exp \left(-\xi  \alpha ^*-\alpha  \xi ^*-2 |\alpha|^2 - |\xi|^2/2   \right)\nonumber\\&+\exp \left(\xi  \alpha ^*+\alpha  \xi ^*-2 |\alpha|^2  - |\xi|^2/2  \right)\Big ]
\end{align}

\begin{align}
\chi_{\alpha_-}(\xi)=&N_{-}^2\Big [\exp\left(\xi  \alpha ^*-\alpha  \xi ^*- |\xi|^2/2  \right)\nonumber\\&+\exp\left(-\xi  \alpha ^*+\alpha  \xi ^*-|\xi|^2/2\right)\nonumber\\
&-\exp \left(-\xi  \alpha ^*-\alpha  \xi ^*-2 |\alpha|^2 - |\xi|^2/2   \right)\nonumber\\&-\exp \left(\xi  \alpha ^*+\alpha  \xi ^*-2 |\alpha|^2  - |\xi|^2/2  \right)\Big ]
\end{align}
where $N_{\pm}^2=\frac{1}{2\pm 2e^{-2|\alpha|^2}}$. 

\subsection*{Performance Benchmark}
Typically the performance of amplifiers can be characterised by the fidelity, F, and the probability of success, $P_s$ of the amplifier. The fidelity metric quantifies the closeness of two quantum states. In this case it quantifies the closeness of the output state after the amplification to an ideal output state. If the ideal output state is pure then F has  a simple form given by
\begin{equation}
\mbox{F}=\tr[\rho_1 \rho_2]
\end{equation}
where $\rho_1,\rho_2$ are some general single mode states and at least one of them is pure. In terms of characteristic functions, as we have already seen, the fidelity is given by
\begin{equation}
\mbox{F}=\int \frac{d^2\xi}{\pi} \chi_1(\xi)\chi_2(-\xi)
\end{equation} 
The probability of success metric on the other hand depends on the working details of the amplifier. For the SCAMP the probability of success is defined as the joint probability of the second detector clicking given that the first detector did not register a click, $P_s=P(D_1=\xmark,D_2=\checkmark)$. The probability for a detector on Geiger mode to not register a click is given by the Kelley-Kleiner formula \cite{Kelley}
\begin{equation}
P_\xmark=\tr[\rho \hat{\pi}_\xmark]=\tr[\rho:\exp(-\eta \hat{a}^\dagger\hat{a}):]
\end{equation}
where $\eta$ is the quantum efficiency of the detector and the colons indicate normal ordering of the operators. In terms of the characteristic function the no counts operator is given by 
\begin{equation}
\chi_{\hat{\pi}_\xmark}(\xi)=\frac{1}{\eta}\exp(-\frac{2-\eta}{2\eta}|\xi|^2)
\end{equation}
and the probability of success is given by
\begin{equation}
P_\xmark=\int \frac{d^2\xi}{\pi }\chi(\xi)\chi_{\hat{\pi}_\xmark}(\xi)
\end{equation}
The probability of success if a detector clicks on the other hand is give by
\begin{equation}
P_\checkmark=\tr[\rho \hat{\pi}_\checkmark]=\tr[\rho:\mathds{1}-\exp(-\eta\hat{a}^\dagger \hat{a}):]
\end{equation}
The click operator, $\hat{\pi}_\checkmark$, in the characteristic function formalism is given by
\begin{equation}
x_{\hat{\pi}_\checkmark}(\xi)=\pi\delta^{(2)}(\xi)-x_{\hat{\pi}_\xmark}(\xi)
\end{equation}
and the probability of success is given by
\begin{equation}
P_\checkmark=\int \frac{d^2\xi}{\pi}\chi(\xi)\chi_{\hat{\pi}_\checkmark}(\xi)
\end{equation}
\subsection*{Output of Comparison Stage}
Now that we have introduced the formalism of the characteristic functions we can have a close look at what happens when a coherent state $\ket{\alpha}$ and a squeezed vacuum $\ket{\zeta}$ are incident on a beamsplitter and one of the two output modes is projected onto the vacuum state.

The  input to the beamsplitter is given by
\begin{align}
    \chi_{in}(\xi_1,\xi_2)=\chi_\alpha(\xi_1)\chi_\zeta(\xi_2)
\end{align}
The output (global) state after the beamsplitter can be found by transforming the arguments of the characteristic function of the input as follows
\begin{align}
    \chi_{out}(\xi_3,\xi_4)=\chi_\alpha(t_1 \xi_3+r_1 \xi_4)\chi_\zeta(t_1\xi_4-r_1 \xi_3)
\end{align}
Postselection then on zero clicks, assuming a perfect detector, leads to
\begin{multline}
    \chi_4(\xi_4)=\frac{1}{\pi P_\xmark}\int \chi_{out}(\xi_3,\xi_4) \chi_{\hat{\pi}_\xmark}(\xi_3)d^2\xi_3\\=\exp\Bigg[-\frac{x^2\{\cosh{s}+(1-r_{1}^2)\sinh{s}\}}{2(\cosh{s}-(1-r_{1}^2)^2\sinh{s}^2)}\\
    -\frac{y^2\{\cosh{s}-(1-r_{1}^2)\sinh{s}\}}{2(\cosh{s}-(1-r_{1}^2)^2\sinh{s}^2)}\\
    -2 i r_1 x Im[\alpha]\cosh{s} \frac{(\cosh{s}+(1-r_{1}^2)\sinh{s})}{\cosh{s}^2-(1-r_{1}^2)^2\sinh{s}^2}\\
     +2 i r_1 y Re[\alpha]\cosh{s} \frac{(\cosh{s}-(1-r_{1}^2)\sinh{s})}{\cosh{s}^2-(1-r_{1}^2)^2\sinh{s}^2}\Bigg]
     \label{chi1}
\end{multline}
where $P_\xmark$ is given by
\begin{multline}
    P_\xmark=\frac{1}{\pi}\iint  \chi_{out}(\xi_3,\xi_4) \chi_{\hat{\pi}_\xmark}(\xi_3)d^2\xi_3d^2 \xi_4\\
    =\sqrt{\frac{1}{(e^{-s}+r_{1}^2\sinh{s})(e^s-r_{1}^2\sinh{s})}}*\\
    \exp\Bigg[-\frac{e^s (1-r_{1}^2)\sinh{s}\mbox{Re}[\alpha]^2}{e^s-r_{1}^2\sinh{s}}\\
    -\frac{e^{-s}(1-r_{1}^2)\sinh{s}\mbox{Im}[\alpha]^2}{e^{-s}+r_{1}^2\sinh{s}}\Bigg]
\end{multline}

The characteristic function of a squeezed coherent state, $\hat{S}(\zeta)\ket{\alpha}$ is given by

\begin{multline}
\chi_{\zeta,\alpha}(\xi)=\exp\big(-x^2 e^{2s}/2-y^2 e^{-2s}/2\\
-2i e^s x \mbox{Im}[\alpha]+2i e^{-s} \mbox{Re}[\alpha]\big)
\label{chi2}
\end{multline}
 where $\xi=x+iy$.
 
 If we compare Eq. \ref{chi1} to Eq.\ref{chi2} we can see that they both represent the same state with the transformations $s \rightarrow s^\prime$, and $\alpha \rightarrow\alpha^\prime$ given by
 
 \begin{align}
     s^\prime&=\ln{\sqrt{\frac{\cosh{s}+(1-r_{1}^2)\sinh{s}}{\cosh{s}-(1-r_{1}^2)\sinh{s}}}}\\
 \alpha^\prime&=\frac{r_1 \alpha \cosh{s}}{\sqrt{(\cosh{s})^2-(1-r_{1}^2)^2 (\sinh{s})^2}}
 \end{align}
Then we can think of the comparison stage as a quantum channel that transforms a coherent state to a squeezed coherent state of squeezing parameter $s^\prime$ and scaled coherent amplitude $\alpha^\prime$. The linearity of quantum mechanics then implies that a linear superposition of coherent states, such as cat states, are transformed in a similar manner. Therefore, the comparison stage transforms cat states into squeezed cat states of squeezing parameter $s^\prime$ and cat size $\alpha^\prime$.

\end{document}